\documentstyle[eqsecnum,aps,twocolumn]{revtex}

\begin{document}
\draft
\title{
Two-loop calculation of the scaling behavior of two-dimensional forced
Navier-Stokes equation
}
\author{J. Honkonen$^a$, Yu. S. Kabrits$^b$ and  M. V. Kompaniets$^b$}
\address{$^a$Theoretical Physics Division, Department of Physical Sciences, P.O.Box
64, 00014 University of Helsinki, Finland
}
\address{$^b$Department of Theoretical Physics, St Petersburg University,
Ulyanovskaja 1, St Petersburg, Petrodvorez, 198904 Russia.}
\date{\today}
\maketitle
\begin{abstract}
Asymptotic properties of the solution of
two-dimensional randomly forced Navier-Stokes
equation with long-range correlations of the driving force are
analyzed
in the two-loop order of perturbation theory
with the use of renormalization group.
Kolmogorov constant of the energy spectrum is calculated
for both the inverse energy cascade and the direct enstrophy
cascade in the second order of the $\varepsilon$ expansion.
\end{abstract}

\narrowtext

\section{INTRODUCTION}

Asymptotic properties of the stochastic Navier-Stokes equation
have been actively studied with the aid of renormalization group
(RG) during the last two decades (for a comprehensive review,
see~\cite{Adzhemyan99}). A random force
with powerlike falloff of spatial correlations
has been used to maintain a stationary scale-invariant regime with
a subsequent generation of an $\varepsilon$ expansion of
scaling exponents and scaling functions.

Most work has been carried out in view of application to
three-dimensional turbulence; partly because of greater physical
interest than in the two-dimensional case, partly due to
additional divergences, which occur in two dimensions and prevent
a direct use of the previously obtained general $d$-dimensional results.
Moreover, there were some serious flaws in the early
work~\cite{Olla91} on the RG-analysis of the forced Navier-Stokes
equation in two dimensions, and only recently has a consistent
renormalization procedure been put forward~\cite{Honkonen96,Honkonen98}
and used in various problems involving randomly forced
incompressible fluid~\cite{Antonov97,Hnatich99,Hnatich00}.

Apart from energy, in two dimensions the enstrophy (squared
vorticity) is an inviscid conserved quantity quadratic in the
velocity. Therefore, two self-similar regimes
corresponding to an inverse energy cascade
towards small wave numbers and a direct enstrophy cascade towards large wave numbers
are expected to take place~\cite{Kraichnan67} instead of
the direct energy cascade observed in three-dimensional turbulence.
The energy (enstrophy) pumping leading to a steady
state with the two scaling regimes may be realized in two
different ways. On one hand, in numerical simulations~\cite{Maltrud91} and some
experiments~\cite{Sommeria86} the energy and enstrophy pumping takes
place on scales in between the inverse energy cascade and the
enstrophy cascade.
On the other hand, in atmospheric turbulence~\cite{Larsen82} the
energy and enstrophy sources are at the outer edges of the scaling
intervals, and it is not clear whether there is an energy and
enstrophy sink between them~\cite{Larsen82} or they coexist~\cite{Maltrud91}.
In both cases the Kolmogorov spectrum of the inverse energy
cascade $E(k)\propto k^{-5/3}$ for $k\ll k_I$ is observed
experimentally and in the majority of simulations for wave numbers smaller than the
characteristic wave number $k_I$ of
the energy pumping.
For the enstrophy inertial range the existing data is not so clear-cut,
although recent high-precision numerical simulations~\cite{Lindborg00} and accurate
analysis of the atmospheric data~\cite{Lindborg99} seem to support the
existence of an energy spectrum $\propto k^{-3}$ in the enstrophy
cascade.

In this paper we have carried out a two-loop renormalization of
the field theory generated by the two-dimensional randomly
forced Navier-Stokes equation. We have taken into account the
divergences specific of two dimensions in the way proposed in
Ref.~\cite{Honkonen96}, and calculated the asymptotic expression
for the energy spectrum in the second order of the $\varepsilon$
expansion for both the energy and enstrophy inertial ranges. For
convenience of calculations, we have used the stochastic
Navier-Stokes equation with a subsequent dimensional
regularization of divergences in two dimensions~\cite{Honkonen96}
instead of the specifically two-dimensional setup~\cite{Honkonen98} with the
stream-function description of the fluid flow.

\section{RENORMALIZED FIELD THEORY FOR THE TWO-DIMENSIONAL STOCHASTIC NAVIER-STOKES EQUATION}

Consider the stochastic Navier-Stokes equation for the
flow of homogeneous incompressible fluid, which for the transverse components of
the velocity field assumes the form
\begin{equation}
\label{1}
\partial _tv_i+P_{ij}v_l\partial_lv _j=\nu _0 \nabla^2 v_i-\xi_0v_i+F_i\,,
\end{equation}
together with the incompressibility condition $\partial _iv _i=0$.
In Eq. (\ref{1}) $v _i(t,{\bf x})$ are the coordinates of the
divergenceless velocity  field, $\nu_0 $ is the kinematic
viscosity, $\xi_0$ is the coefficient of friction, and
$P_{ij}$ is the transverse projection operator ($P_{ij}=\delta _{ij}
-k_ik_j/k^2$ in the wave-number space), and $F_i$ are the coordinates of
the random force. Here, and henceforth, summation over
repeated indices is implied.

In experimental realizations and simulations of a
two-dimensional turbulent flow energy may be consumed not only by microscale
dissipation, but also by the friction at the boundaries of the fluid
layer. The friction term in Eq. (\ref{1}) makes it possible to maintain stationary state with
the anticipated inverse energy cascade towards small wave numbers and the direct enstrophy
cascade towards large wave numbers, when the
pumping is carried out in between the corresponding inertial
ranges. The coefficient of friction is a mass term from the
point of view of renormalization, therefore we put $\xi_0=0$ in
the calculation of the renormalization constants of the solution of Eq. (\ref{1}).
As shown in Ref.~\cite{Honkonen98}, the friction term in Eq. (\ref{1}) is not
renormalized and thus does not affect the RG equations and the
subsequent asymptotic analysis.

In the applications of the stochastic
Navier-Stokes equation~(\ref{1}) to turbulence
the random force is assumed to
have a gaussian distribution with zero mean and the
correlation function in the wave-vector space~\cite{Adzhemyan99} of the form
\begin{equation}
\label{2}
\langle F_i(t,{\bf k})F_j(t',{\bf k}')\rangle=P_{ij}(2\pi)^{d/2}\delta ({\bf k}+{\bf k}')
\delta (t-t')d_F(k)\,.
\end{equation}
The scalar kernel has a powerlike asymptotic behavior at large
wave numbers:
\begin{equation}
\label{correlator0}
d_F(k)=D_0k^{4-d-2\varepsilon}h(m/k)\,,
\end{equation}
where $h(x)$ is a well-behaved function of the dimensionless
argument $m/k$ ensuring the convergence of the inverse Fourier
transform of $d_F(k)$ at small $k$ and with the large $k$ behavior
fixed by the condition $h(0)=1$. In a fixed dimension above two
dimensions the force correlation function is not renormalized and
the the kernel (\ref{correlator0}) remains intact. This is,
however, not the case in two dimensions, in which renormalization
generates additional terms $\propto k^2$ into the force
correlation function. In order to deal with a multiplicatively
renormalizable theory -- which is convenient technically -- we add
this term to the correlation kernel at the outset and use, instead
of the function (\ref{correlator0}), the modified function
\begin{equation}
\label{d_F}
d'_F(k)=D_{01}k^{4-d-2\varepsilon}h(m/k)+D_{02}k^2\,.
\end{equation}
The force correlation function is related to two
basic physical quantities, the energy pumping rate ${\cal E}$ and the enstrophy
pumping rate ${\cal B}$, as
\begin{eqnarray}
\label{EB}
{\cal E}=\frac{d-1}{2}\int \frac{d{\bf
k}}{(2\pi)^d}d'_F(k)\,,\nonumber\\
{\cal B}=\frac{d-1}{2}\int \frac{d{\bf k}}{(2\pi)^d}k^2 d'_F(k)
\end{eqnarray}
in $d$-dimensional space, which allows to connect the "coupling constant" $D_{01}$ with
the pumping rate in the corresponding asymptotic region.

We cast the stochastic problem~(\ref{1}), (\ref{2}), (\ref{d_F})
into a field theory with the De-Dominicis-Janssen "action" in the usual manner~\cite{Adzhemyan99}.
An analysis of UV
divergences with the use of Galilei invariance, causality and
symmetries of the model allows to write the renormalized action in
the form
\begin{eqnarray}
\label{Raction}
S=
{1\over 2}\int\!{dtd{\bf k}\over (2\pi)^d}\,\tilde{{\bf v}}\nonumber\\
\times\Biggl[g_1\nu^3\mu^{2\varepsilon}
(k^2)^{1-\delta-\varepsilon}h(m/k)+g_2\nu^3\mu^{-2\delta}Z_2k^2\Biggr]\tilde{{\bf v}}\\
+\int\!dtd{\bf x}\,\tilde{{\bf v}}\cdot\left[\partial_t {\bf v} +
 ({\bf v} \cdot {\nabla} ) {\bf v} - \nu Z_1\nabla^2 {\bf v}\right]\nonumber\,,
\end{eqnarray}
where $\mu$ is the scale-setting parameter of the renormalized
model and $2\delta=d-2$ is the parameter of
dimensional regularization. Only two renormalization constants
$Z_1$ and $Z_2$ are needed to absorb the UV divergences of the
model in two dimensions. To avoid excessive notation,
we have used the same symbols for both the
fields and their Fourier transforms in (\ref{Raction}).

Renormalized parameters of the action (\ref{Raction}) are defined by
\begin{eqnarray}
\label{renormalized}
\nu_0&=&\nu Z_{1}\,,\\
D_{01}&=&g_1\nu^3\mu^{2\varepsilon}Z_1^{-3}\,,\\
D_{02}&=&g_2\nu^3\mu^{-2\delta}Z_2Z_1^{-3}\,.
\end{eqnarray}
We have used a combination of dimensional and analytic regularization with
the parameters $\varepsilon$ and $2\delta=d-2$. As a consequence, the UV divergences appear as
poles in linear combinations of the regularizing parameters.
Normalization has been fixed by the choice of the minimal subtraction (MS)
scheme~\cite{Zinn-Justin89}.

\section{RENORMALIZATION-GROUP EQUATIONS AND FIXED POINTS}

We set up the notation and basic equations for
the spatial Fourier transform of the pair
correlation function of the random velocity field
 \begin{eqnarray}
\label{correlator}
 W_{mn}(t_1-t_2,{\bf k}; D_{01},D_{02},\nu_0)=\int\,
 \frac{\mbox{d}^d{\bf x}_1}{(2\pi)^{d}}\,\nonumber\\
\times \langle\,v_m(t_1,{\bf x}_1)\,v_n(t_2,{\bf x}_2)\,\rangle
 \,e^{ i {\bf k} \cdot ({\bf x}_1-{\bf x}_2)}\,,
 \end{eqnarray}
because this quantity is directly connected to the energy spectrum
through the relation
$\langle v_n(t,{\bf x}) v_n(t,{\bf x})\rangle=2\int_0^\infty E(k)dk$.

Independence of the unrenormalized
pair correlation
function
\[
W_{mn}(t,{\bf k}; D_{01},D_{02},\nu_0)=W_{mn}^R(t,{\bf
k};g_1,g_2,\nu,\mu)
\]
of the velocity field ${\bf v}$ of the
scale-setting parameter $\mu$ gives rise to the
basic RG equation
\begin{equation}
\label{RG}
 \Biggl[\,{\mu}
 {\partial\over\partial \mu} + {\beta}_{1}{\partial\over\partial g_{1}} +
 {\beta}_{2}{\partial\over\partial g_ {2}}
-\gamma_1\nu{\partial\over\partial \nu}\,\Biggr]\,
W^{R}_{ mn}=0
\end{equation}
for the renormalized correlation function $W_{mn}^R$.
The coefficient functions of
Eq.~(\ref{RG}) $\beta_{1}$, $\beta_{2}$, and $\gamma_1$ are expressed in terms of
logarithmic derivatives
of the renormalization constants. We use the definitions
 \begin{equation}
 \gamma_i=\mu \frac{\partial \ln Z_i}{\partial \mu}\Biggl|_{0}\,,\,\,
\,\,\,
 \beta_i=\mu \frac{\partial g_i}{\partial\mu }\Biggl|_{0}\,,
 \,\,
 \label{defbeta}
 \end{equation}
where $ i=1,2$,
and the subscript "0" refers to partial derivatives
taken at fixed values of the bare parameters.

It is convenient to express the correlation function through a dimensionless scalar
function $R$. In two-dimensional space we define this function through the relation
\begin{equation}
\label{Rfunctions}
W_{mn}^{R}(t,{\bf k};g_1,g_2,\nu)={1\over 2}g_1\nu^2P_{mn}({\bf k})R(\tau,s;g_1,g_2)\,,
\end{equation}
where
$s=k/\mu$ is the dimensionless wave number, and $\tau=t\nu k^2$
the dimensionless time.
Solving Eqs. (\ref{RG}),  (\ref{defbeta}) by the method of characteristics we obtain
the correlation function in the form
\begin{equation}
\label{Rsolution}
W_{mn}^{R}(t,{\bf k};g_1,g_2,\nu) =
 {1\over 2}\overline{g}_1\overline{\nu}^2P_{mn}({\bf k})\,
R\left(tk^2\overline{\nu},1; {\overline g}_1,{\overline g}_2\right)\,,
\end{equation}
where  ${\overline g}_i$ are the solution of the Gell-Mann-Low equations:
\begin{equation}
\label{Gell}
\frac{d\overline{g}_i}{d\ln s}= \beta_i
\left[ \overline{g}_1,\overline{g}_2\right]\,,\qquad i=1,2\,,
\end{equation}
and $\overline{\nu}$ is the running coefficient of viscosity
\begin{equation}
\label{viscosity}
\overline{\nu}=\nu e^{-\int_1^s\!{\rm d}x\,\gamma_{1}({\overline
g}_1(x),{\overline
g}_2(x))/x}\,.
\end{equation}
Writing the latter
in terms of the unrenormalized (physical) parameters and the
running coupling constant $\overline{g}_1$~\cite{Adzhemyan99} as
\begin{equation}
\label{scalingfactor}
\overline{\nu}
=\left(\frac{D_{01}}{{\overline
g}_{1}}\right)^{1/3}k^{-2\varepsilon/3}\,,
\end{equation}
we arrive at the expression
\begin{eqnarray}
\label{physical}
W_{mn}^{R}(t,{\bf k};g_1,g_2,\nu)={1\over 2}\,\overline{g}_1^{1/3}D_{01}^{2/3}
k^{-4\varepsilon/3}\nonumber\\
\times P_{mn}({\bf
k})R\left((D_{01}/\overline{g}_1)^{1/3}k^{2-2\varepsilon/3}t,1;
\overline{g}_1,\overline{g}_2\right)\,.
\end{eqnarray}
For the $\beta $-functions:
\begin{eqnarray}
\label{2loopbeta}
\beta _{1}&=&g_1(-2\varepsilon +3\gamma_1)\,,\nonumber\\
\beta _{2}&=&g_2(-\gamma_2+3\gamma_1)\,,
\end{eqnarray}
a tedious two-loop calculation, with the use of the step function
$h(m/k)=\theta(k-m)$ in the kernel (\ref{d_F}), and dimensional regularization,
 yields
\begin{eqnarray}
\label{2loopgamma}
\gamma_1=u_1+u_2+u_1^2\left({1\over 2}\alpha+{3\over 4}-r\right)\nonumber\\
+u_1u_2\left(\alpha+{3\over 2}-2r\right)+u_2^2\left(-{1\over
2}-r\right)\nonumber\\
\gamma_2={(u_1+u_2)^2\over u_2}+{u_1^3(1-r)\over u_2}\\
+u_1^2\left({5\over 2}\alpha+{31\over 4}-3r\right)\nonumber\\
+u_1u_2\left({5\over 2}\alpha+{25\over 4}-3r\right)+u_2^2\left(-{1\over
2}-r\right)\,,\nonumber
\end{eqnarray}
where
\begin{equation}
u_1={g_1\over 32\pi}\,,\qquad u_2={g_2\over 32\pi}\,.
\end{equation}
The constant $r=-0.1685$ in Eq. (\ref{2loopgamma}) comes from a numerical calculation of
those parts of two-loop graphs, for which analytic results were not
feasible. The constant $\alpha=C-\ln(4\pi)=-1.9538$ ($C$ is Euler's constant) is brought
about by a $2\delta=d-2$ expansion of the geometric factor
$S_d/(2\pi)^d=2/[\Gamma(d/2)(4\pi)^{d/2}]={1\over
2\pi}[1+\alpha\delta+O(\delta^2)]$. The method of calculation is
essentially the same as was used for the $d$-dimensional case in
Ref.~\cite{Adzhemyan00}

At one-loop order the $\gamma$-functions (\ref{2loopgamma})
are exactly the same as those of the $d$-dimensional
Navier-Stokes equation in two dimensions~\cite{Honkonen96}.
They also coincide with the expressions obtained directly in two dimensions for the corresponding
stochastic vorticity equation~\cite{Honkonen98}. Thus,
we think that for calculation of the coefficient functions of the renormalization group equation
the results of the $d$-dimensional model
in the two-parameter expansion
may be applied directly to the two-dimensional case. There might
be some discrepancies in calculations involving composite
operators due to different symmetries in two-dimensional and
general $d$-dimensional cases, but we do no calculate anything
like that here.

The fixed points are determined by the system of equations
$\beta_1=\beta_2=0$.
From the solution of Eqs. (\ref{Gell}) near a fixed point
it follows that the fixed point is infrared
stable, when the matrix $\omega_{nm} =\partial _n\beta _m$ is positive
definite. If  $\varepsilon <0$, then
the trivial fixed point: $u_1^*=u_2^*=0$ is infrared
stable.
The anomalous asymptotic behavior of the model at small wave numbers
is governed by the nontrivial fixed point
\begin{eqnarray}
\label{fixedpoint}
u_1^*=\frac{4}{9}\varepsilon-{2\over 27}(2\alpha+5-4r)\,,\nonumber\\
u_2^*=\frac{2}{9}\varepsilon-{4\over 81}(\alpha-2-3r)\,,
\end{eqnarray}
at which the eigenvalues of the stability matrix are
\begin{eqnarray}
\label{eigen}
\omega _{1,2}=\left({4\over 3}\pm i{2\sqrt{2}\over 3}\right)\varepsilon\nonumber\\
+
\left(-{2\over 3}-{4\over 9}r\pm i{3-2r\over 9}\,\sqrt{2}\right)\varepsilon^2\,.
\end{eqnarray}
Note that these eigenvalues are independent of $\alpha$.
The real parts of both eigenvalues (\ref{eigen}) are positive and the fixed point
(\ref{fixedpoint}) infrared stable, when
$\varepsilon>0$.

\section{TWO-LOOP CALCULATION OF KOLMOGOROV CONSTANTS}

The connection between the energy spectrum $E(k)$
and the equal-time correlation function of the velocity field,
$\langle v_n({\bf x}) v_n({\bf x})\rangle=2\int_0^\infty E(k)dk$,
in the
two-dimensional wave-vector space amounts to
\begin{equation}
\label{connection1}
E(k)=\frac{k}{4\pi}W_{nn}(0,{\bf k})\,.
\end{equation}
From (\ref{physical}) the asymptotic expression
\begin{equation}
\label{asyspectrum}
E(k)=g_*^{1/3}{D_{01}}^{2/3}\frac{k^{1-4\varepsilon/3}}{8\pi}\,
R\left(0,1;g_1^*,g_2^*\right)
\end{equation}
follows, when $k\to 0$. Here,
$g_1^*=32\pi u_1^*$, $g_2^*=32\pi u_2^*$ are the values of the coupling
constants at the infrared-stable fixed point~(\ref{fixedpoint}).

The relations (\ref{EB}) allow to express the parameters $D_{01}$ and $D_{02}$
in terms of the energy (or enstrophy) pumping rates ${\cal E}$ (${\cal
B}$). Integrating over the wave-number shell $m<k<\Lambda$ with
$h(x)=1$ in the kernel (\ref{d_F}) we obtain, in the limit of
widely separated upper and lower wave-number limits,
\begin{eqnarray}
\label{Eg}
{\cal E}=\frac{D_{01}}{8\pi}{\Lambda^{2(2-\varepsilon)}\over
2-\varepsilon}
+\frac{D_{02}}{16\pi}\Lambda^{4}\,,\\
\label{E3}
{\cal B}=\frac{D_{01}}{8\pi}{\Lambda^{2(3-\varepsilon)}\over
3-\varepsilon}
+\frac{D_{02}}{24\pi}\Lambda^6\,.
\end{eqnarray}
The spectrum~(\ref{asyspectrum}) should be independent of the details of the
energy pumping, i.e. independent of  the upper cutoff
$\Lambda$ in the range $m\ll k\ll \Lambda$. According to the relation (\ref{Eg}),
this goal is achieved by the choice $D_{02}=0$ and
$\varepsilon=2$ for the anticipated inverse energy cascade.
The relation (\ref{E3}), in turn, shows that
the choice $D_{02}=0$ and
$\varepsilon=3$ leads to scale-invariant behavior for the direct
enstrophy cascade.  In both cases it should be borne in mind that
the bare coupling constant $D_{02}$ is technically a book-keeping parameter reflecting the
necessity of the introduction of the short-range term in the correlation function of
the random force. Physically, it could be related to the intensity
of thermal fluctuations, which, however,  are irrelevant in the
energy balance
of stationary developed turbulence.

The Kolmogorov constants are determined from the asymptotic relations
\begin{eqnarray}
\label{defCK}
E(k)=C(\varepsilon){\cal E}^{2/3}k^{-5/3}\left({\Lambda\over
k}\right)^{4(\varepsilon-2)/3}\,,\\
\label{defCK'}
E(k)=C'(\varepsilon){\cal B}^{2/3}k^{-3}\left({\Lambda\over
k}\right)^{4(\varepsilon-3)/3}\,.
\end{eqnarray}
Thus, the choice $\varepsilon=2$ in (\ref{Eg}) renders the spectrum~(\ref{asyspectrum})
completely scale-invariant with the Kolmogorov exponents
corresponding to the energy cascade, whereas the substitution $\varepsilon=3$ in
(\ref{E3}) leads to scale-invariant behavior in the enstrophy cascade.

The Kolmogorov constants may be calculated in the $\varepsilon$ expansion
from (\ref{asyspectrum}), (\ref{defCK}) and  (\ref{defCK'}).
The present two-loop calculation allows to find
correction terms to the previosly found~\cite{Honkonen98}
expressions. The result is
\begin{eqnarray}
\label{CK}
C(\varepsilon)= 2\cdot 3^{1/3}\varepsilon^{1/3}\left[1+{2\over
9}(1+r)\varepsilon\right]\,,\\
\label{CK'}
C'(\varepsilon)= 3\cdot 2^{1/3}\varepsilon^{1/3}\left[1+{3+2r\over
9}\varepsilon\right]\,.
\end{eqnarray}
For $\varepsilon=2$ we obtain from (\ref{CK})
$C=4.977$.
The closure model leads to the prediction $C=6.69$~\cite{Kraichnan67}. Results of numerical
simulations vary from $C=2.9$~\cite{Herring85} to $C\sim 14$~\cite{Siggia81}.
Experimental results~\cite{Sommeria86} yield the range $3<C<7$.
The result obtained here is thus in better agreement with other
available data than the leading-order value of the $\varepsilon$
expansion
$C=3.634$ obtained in Ref.~\cite{Honkonen98}.

For $\varepsilon=3$ the value
$C'=10.29$  obtained from (\ref{CK'})
is significantly larger than the leading-order value
$C'=5.451$ of Ref.~\cite{Honkonen98} and the discrepancy between
the present
result and the closure-model prediction
$C'=2.626$~\cite{Kraichnan67} is larger. However, from a more detailed analysis of the model
it may be concluded that calculation of
the constant $C'$ is not unambigous. The
value obtained from (\ref{CK'}) corresponds to the case, in which
the coefficient of friction $\xi_0=0$. If, however, this
coefficient is retained, then the asymptotic expression for the
energy spectrum is~\cite{Honkonen98}
\begin{eqnarray}
\label{asyspectrumxi}
E(k)=g_*^{1/3}{D_{01}}^{2/3}\frac{k^{1-4\varepsilon/3}}{8\pi}\nonumber\\
\times
R\left(0,1;g_1^*,g_2^*,\frac{(g_1^*)^{1/3}\xi_0}{D_{01}^{1/3}}
k^{-(2-2\varepsilon/3)}\right)
\end{eqnarray}
instead of (\ref{asyspectrum}). The energy spectrum
(\ref{asyspectrumxi}) is scale-invariant for $\varepsilon=3$
regardless of the value $\xi_0$. Therefore,
the proportionality constant in the scaling law seems to be
nonuniversal in the enstrophy inertial range,
and depends on the properties of large-scale dissipation~\cite{Honkonen98}.
Recent spectral closure analysis and numerical simulations have led to
similar conclusions~\cite{Kaneda01}.

\section{CONCLUSION}

In this paper
we have carried out a two-loop renormalization of the randomly forced
Navier-Stokes
equation with long-range correlated random force in two dimensions
in view of two
different patterns of scale-invariant asymptotic behavior.

We have calculated the Kolmogorov constant for a powerlike asymptotic energy
spectrum $\propto k^{-5/3}$ of the random velocity field in the inertial
range of the inverse energy cascade in the second order of an $\varepsilon$ expansion
with the result $C=4.977$, which is in reasonable agreement with
other
available experimental and theoretical data.

We have also calculated the Kolmogorov constant for the
spectrum $\propto k^{-3}$ in the inertial
range of the direct enstrophy cascade with the second-order result
$C'=10.29$ with a larger deviation than at the leading order from results
obtained by other methods. However, explicit asymptotic
expressions for the pair correlation function of the random velocity field
obtained in the present approach strongly indicate that the
Kolmogorov constant in the enstrophy cascade is not universal, but
depends on the enstrophy dissipation due to large-scale friction.

\acknowledgments
This work was supported in part by the Nordic Grant for Network Cooperation with the Baltic
Countries and Northwest Russia No.~FIN-18/2001.

\end{document}